\documentclass[prb,twocolumn,showpacs,preprintnumbers,amsmath,amssymb]{revtex4}
\usepackage{graphicx}

\begin{document}

\title{Theoretical study of angle-resolved two-photon photoemission in two-dimensional insulating cuprates}

\author{H. Onodera}
\author{T. Tohyama}
 \altaffiliation[Corresponding author: ]{tohyama@imr.tohoku.ac.jp}
\author{K. Tsutsui}
\author{S. Maekawa}
 \altaffiliation[Also at ]{CREST, Japan Science and Technology Agency (JST), Kawaguchi, Saitama 332-0012, Japan.}
\affiliation{Institute for Materials Research, Tohoku University, Sendai, 980-8577, Japan.}
\date{\today}

\begin{abstract}
We propose angle-resolved two-photon photoemission spectroscopy (AR-2PPES) as a technique to detect the location of the bottom of the upper Hubbard band (UHB) in two-dimensional insulating cuprates.  The AR-2PPES spectra are numerically calculated for small Hubbard clusters.  When the pump photon excites an electron from the lower Hubbard band, the bottom of the UHB is less clear, but when an electron in the nonbonding oxygen band is excited, the bottom of the UHB can be identified clearly, accompanied with additional spectra originated from the spin-wave excitation at half filling.
\end{abstract}

%\keywords{}%Use show keys class option if keyword display desired

\maketitle

The charge gap in Mott insulators is a consequence of strong electron correlation.  The nature of excitations across the gap is controlled by both the lower Hubbard band (LHB) and the upper Hubbard band (UHB). Therefore, the clarification of the momentum dependence of the two bands will be very crucial for understanding the nature of the Mott-gap excitation.

Parent compounds of high-$T_c$ superconductors such as La$_2$CuO$_4$ and Ca$_2$CuO$_2$Cl$_2$ are a good example of the Mott insulator in two dimensions (2D).  It was theoretically proposed that the top of the LHB [more precisely the Zhang-Rice singlet band (ZRB) (Ref.~1)] is located at $\mathbf{k}=(\pm\pi/2, \pm\pi/2)$, while the bottom of the UHB is at $(\pi,0)$ and $(0,\pi)$.~\cite{Tsutsui} The location of the top of the LHB has clearly been observed by angle-resolved photoemission experiments.~\cite{Wells} On the other hand, the bottom of UHB has not been directly observed yet, but only indirectly confirmed by resonant inelastic x-ray scattering that reveals momentum-dependent Mott gap excitations.~\cite{Hasan} 

In this Brief Report, we propose angle-resolved two-photon photoemission spectroscopy (AR-2PPES) as a new technique to detect the location of the bottom of the UHB in the 2D insulating cuprates.  The two-photon photoemission spectroscopy has been widely used for the studies of the electronic excitations at metal surfaces and their decay in the time domain.~\cite{Weinelt,Echenique}  On the other hand, for strongly correlated systems such as high-$T_c$ cuprates there are a few studies only in the metallic and superconducting regions of Bi$_2$Sr$_2$CaCu$_2$O$_{8+\delta}$,~\cite{Nessler,Sonoda} but no report on insulating materials.  Therefore, the application of AR-2PPES to the insulating cuprates will open a new way of spectroscopic study for strongly correlated electron systems.

The AR-2PPES spectrum is numerically calculated for small clusters described by the Hubbard model with realistic parameters.  We consider two types of pump-photon excitation: One corresponds to the case where the pump photon induces an excitation from LHB to UHB, while the other is an excitation from the nonbonding (NB) oxygen band in which the oxygen 2$p_\sigma$ orbitals form an asymmetric state with respect to the Cu3$d_{x^2-y^2}$ orbital.  In the former, we find two kinds of AR-2PPES spectra that reflect information on either the UHB or LHB.  However, the location of the bottom of UHB is not clearly identified because of diffusive features in the spectra.  In the latter case, we clearly find the location of the bottom of UHB.  In addition to this, new spectral weights that are not seen in  inverse photoemission spectra emerge in AR-2PPES.  The origin of the additional spectra is attributed to the spin-wave excitation expected at half filling.  These theoretical results will be useful for the analysis of AR-2PPES experimental data in the near future. 

We employ a single-band Hubbard model with long-range hoppings to describe the electronic states in the 2D insulating cuprates.  The Hamiltonian reads
\begin{eqnarray}\label{HtU}
H_\mathrm{Hub} &=& -\sum_{\langle\mathbf{i},\mathbf{j}\rangle,\sigma} t_{ij}\left( c_{\mathbf{i},\sigma}^\dagger c_{\mathbf{j},\sigma}+\mathrm{H.c.} \right) \nonumber \\
  &&+U\sum_\mathbf{i} n_{\mathbf{i},\uparrow}n_{\mathbf{i},\downarrow} - \mu \sum_{\mathbf{i},\sigma} n_{\mathbf{i},\sigma}\;,
\end{eqnarray}
where $c_{\mathbf{i},\sigma}^\dagger$ is the creation operator of an electron
with spin $\sigma$ at site $\mathbf{i}$, and $n_{\mathbf{i},\sigma}=c_{\mathbf{i},\sigma}^\dagger c_{\mathbf{i},\sigma}$. The summations $\langle \mathbf{i},\mathbf{j} \rangle$ run over neighboring pairs up to the third-nearest neighbors. The hopping parameter $t_{ij}$ then contains three terms: $t$, $t'$, and $t''$ for the first, second, and third neighbors.  $U$ is the on-site Coulomb interaction, and $\mu$ is the chemical potential.  We take the parameter values of the $t$-$t'$-$t''$-$U$ model to be $t'/t=-0.34$, $t''/t=0.23$, and $U/t=10$, which are realistic ones for the 2D cuprates.~\cite{Tsutsui}

In AR-2PPES examined in the present work, we consider the situation that the pump photon induces a dipole transition from occupied states to UHB and the probe photon kicks an electron in UHB out of sample.  Generally there are two kinds of contributions for the electron energy distribution of AR-2PPES:~\cite{Shalaev,Ueba} One is due to simultaneous excitations that is given, at zero temperature, by
\begin{eqnarray}\label{Is}
I_\mathrm{s}(E_\mathrm{kin},\omega_1,\omega_2,\mathbf{k})&=&
\frac{2\pi}{N}\sum_{f,\sigma} 
\left|\sum_m\frac{\langle f|c_{\mathbf{k},\sigma}| m\rangle \langle m|j_x|i\rangle}
{E_m-E_i-\omega_1-i\Gamma_{im}}\right|^2 \nonumber \\
&\times& \delta(E_\mathrm{kin}+E_f- E_i-\omega_1-\omega_2)\;,
\end{eqnarray}
and the other is due to two sequential (cascade) excitations, expressed as
\begin{eqnarray}\label{Ic}
&&I_\mathrm{c}(E_\mathrm{kin},\omega_1,\omega_2,\mathbf{k})=
\frac{(2\pi)^2}{N}\sum_{f,\sigma} \sum_m
\frac {\left| \langle f|c_{\mathbf{k},\sigma}| m\rangle
\langle m|j_x|i\rangle \right|^2}
{\Gamma_{mm}} \nonumber \\
&&\ \ \times
\delta(E_m-E_i-\omega_1)
\delta(E_\mathrm{kin}+E_f-E_m-\omega_2)\;,
\end{eqnarray}
where $N$ is the number of sites, $E_\mathrm{kin}$ is the kinetic energy of photoelectron, and $\omega_1$ ($\omega_2$) is the pump-photon (probe-photon) energy.  The state $\left|i\right\rangle$ is the ground state of a half-filled system with energy of $E_i$, and $\left|m\right\rangle$ ($\left|f\right\rangle$) is the intermediate (final) state with energy of $E_m$ ($E_f$), which is the eigenstate of $H_\mathrm{Hub}$.  The current operator $j_x$ connects $\left|i\right\rangle$ with $\left|m\right\rangle$, where an electric field is assumed to have a polarization along the $x$ direction in the square lattice.  We choose the energy relaxation constant $\Gamma_{mm}$ to be twice as large as the phase relaxation constant $\Gamma_{im}$ neglecting the pure dephasing.~\cite{Loudon}  The relaxation constants are also assumed to be independent of $\left| m\right\rangle$: $\Gamma_{mm}=2\Gamma_{im}=2\Gamma$.  We chose $\Gamma=0.2t$.

\begin{figure}
\begin{center}
\includegraphics[width=7cm]{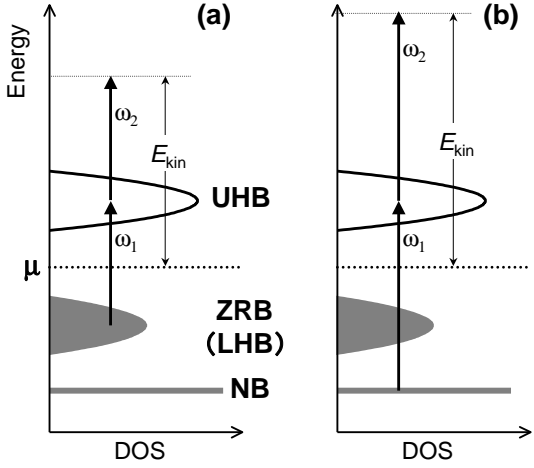}
%\vskip5mm
\caption{\label{fig1}
Schematic picture of two-photon photoemission processes in insulating cuprates.  Two processes are shown: The pump photon with energy $\omega_1$ excites an electron from (a) the lower Hubbard band (LHB) [more precisely the Zhang-Rice singlet band (ZRB)] or (b) the nonbonding (NB) oxygen band to the upper Hubbard band (UHB).  The excited electron is emitted outside with the kinetic energy $E_\mathrm{kin}$ by the probe photon with energy $\omega_2$.}
\end{center}
\end{figure}

In order to calculate $I_\mathrm{s}$ and $I_\mathrm{c}$, we use numerically exact diagonalization techniques for small clusters with periodic boundary conditions.  For $I_\mathrm{s}$ in Eq.~(\ref{Is}), a standard technique combining the conjugate gradient and Lanczos methods is employed. For $I_\mathrm{c}$ in Eq.~(\ref{Ic}), we select some dozens of $\left|m\right\rangle$ that have large values of $|\langle m|j_x|i\rangle|^2$.  The delta functions are broadened by a Lorentzian with a width of $0.2t$.

It is important to notice that AR-2PPES spectrum generally contains two kinds of energy-dependent spectral features.~\cite{Sakaue}  One is the spectrum that satisfies $E_\mathrm{kin}=\omega_1+\omega_2+\epsilon_\mathrm{i}$, $\epsilon_\mathrm{i}$ being the energy level of an occupied state.  Since $E_\mathrm{kin}$ is proportional to $\omega_1+\omega_2$, a peak structure obeying this relation is called the 2$\omega$ peak, from which we can extract knowledge of the occupied state.   The other gives $E_\mathrm{kin}=\omega_2+\epsilon_\mathrm{m}$, $\epsilon_\mathrm{m}$ being the energy level of an unoccupied state.  A peak satisfying this relation is called the $\omega$ peak and provides knowledge of the unoccupied state. 

First, we examine the case where the pump photon induces a dipole transition from ZRB to UHB, as schematically depicted in Fig.~\ref{fig1}(a).  Since ZRB is regarded as the LHB in the single-band Hubbard model, the dipole transition is controlled by a current operator obtained from Eq.~(\ref{HtU}): $j_x=i\sum_{\mathbf{k},\sigma} \alpha_\mathbf{k} c_{\mathbf{k},\sigma}^\dagger c_{\mathbf{k},\sigma}$ with $\alpha_\mathbf{k}=-2t \sin k_x - 4t' \sin k_x \sin k_y + 2t''\sin2k_x$.

\begin{figure}
\begin{center}
\includegraphics[width=7.5cm]{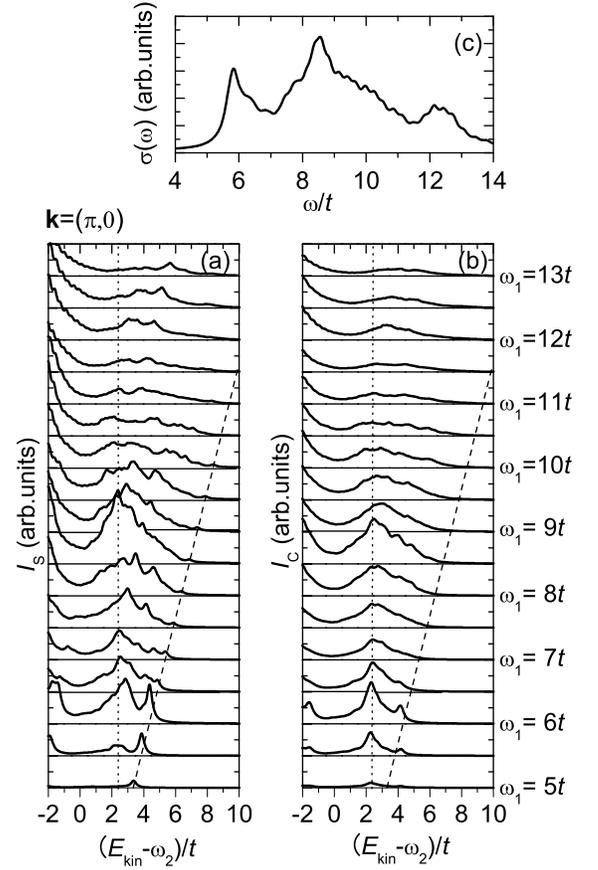}
\caption{\label{fig2}
Pump-photon energy $\omega_1$ dependence of two-photon photoemmision spectra at $\mathbf{k}=(\pi,0)$ in a $4\times 4$ $t$-$t'$-$t''$-$U$ cluster. (a) Simultaneous excitation, and (b) cascade excitation.  The range of $\omega_1$ is determined from the optical conductivity $\sigma(\omega)$ in (c).  The dashed (dotted) line in (a) and (b) denotes a guide to the eyes for the spectrum coming from LHB (UHB).}
\end{center}
\end{figure}

Figures~\ref{fig2}(a) and 2(b) show the dependence of $I_\mathrm{s}$ and $I_\mathrm{c}$, respectively, at $\mathbf{k}=(\pi,0)$ on the pump-photon energy $\omega_1$, calculated by using a half-filled $4\times 4$ cluster where $\mu=4.55t$.  We choose the range of $\omega_1$ ($5t\leqslant\omega_1\leqslant 13t$) from the optical conductivity $\sigma(\omega)$ shown in Fig.~\ref{fig2}(c).  We note that the range examined corresponds to 1.7~eV~$\leqslant\omega_1\leqslant$~4.6~eV, taking $t=0.35$~eV. Since $I_\mathrm{s}$ and $I_\mathrm{c}$ are plotted as a function of $(E_\mathrm{kin}-\omega_2)/t$, the $2\omega$ peak containing information on the LHB appears as a structure whose position increases linearly with $\omega_1$.  Such a peak is seen along the dashed lines.  We find that the $2\omega$ peak in $I_\mathrm{s}$ is clearer than that in $I_\mathrm{c}$. This is due to the fact that $I_\mathrm{s}$ includes a virtual excitation via $\left|m\right\rangle$ but not in $I_\mathrm{c}$.  As for a structure whose position is independent of $\omega_1$, we can find a broad peak at around $E_\mathrm{kin}-\omega_2 = 2.5t$ in both $I_\mathrm{s}$ and $I_\mathrm{c}$.  The weight is enhanced at around $E_\mathrm{kin}-\omega_2 = 8.5t$, where $\sigma(\omega)$ shows a peak.  For the $\omega$ peak, $I_\mathrm{s}$ and $I_\mathrm{c}$ show qualitatively similar $\omega_1$ dependence. 

\begin{figure}
\begin{center}
\includegraphics[width=7.5cm]{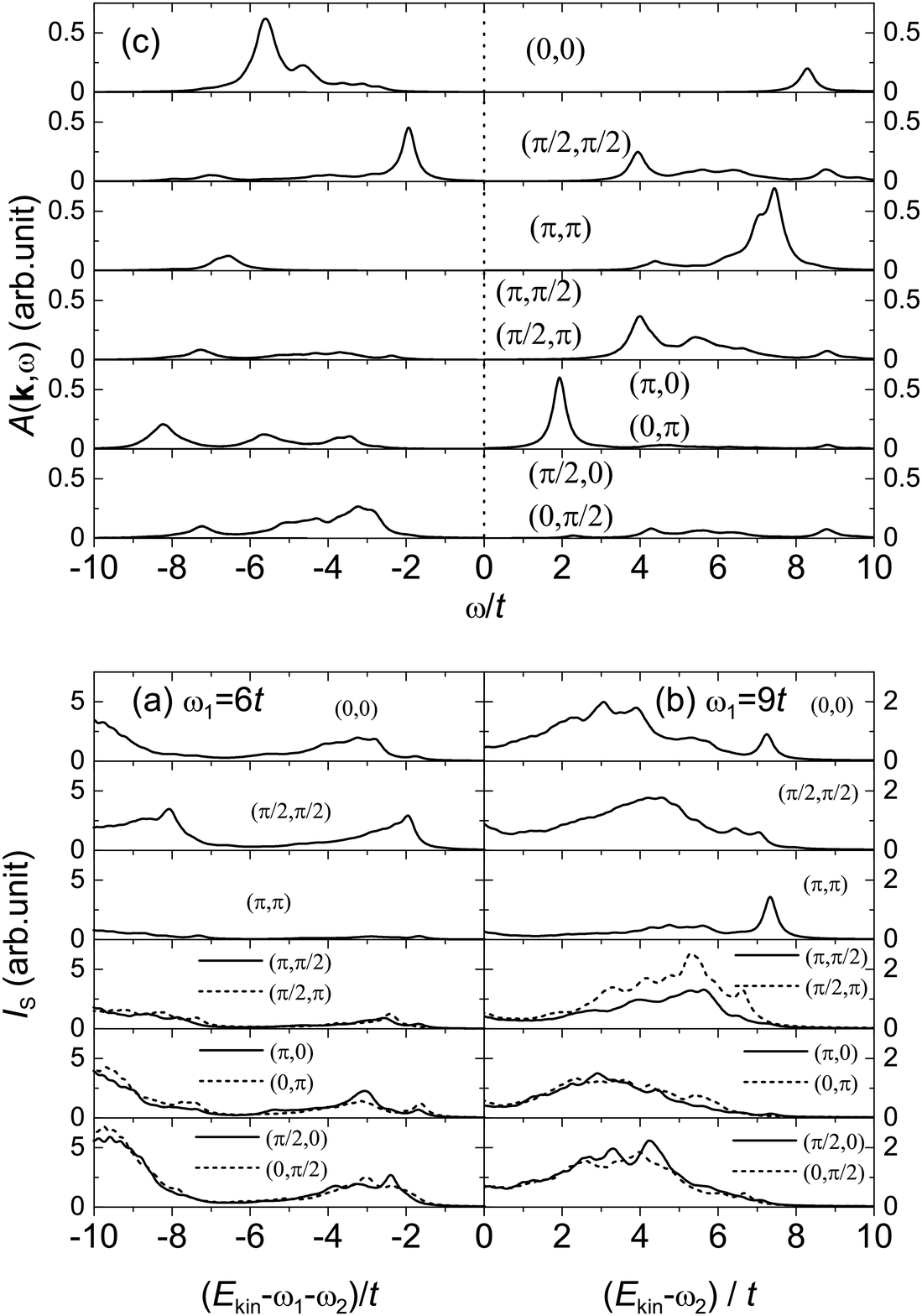}
\caption{\label{fig3}
Momentum dependence of two-photon photoemmision spectra $I_\mathrm{s}$ from simultaneous process in a $4\times 4$ $t$-$t'$-$t''$-$U$ cluster.  The pump-photon energy is that (a) $\omega_1=6t$ and (b) $\omega_1=9t$.  The single-particle spectral function $A(\mathbf{k},\omega)$ at half filling is shown in (c). $A(\mathbf{k},\omega)$ below and above $\omega/t=0$ can be comparable to $I_\mathrm{s}$ in (a) and (b), respectively.}
\end{center}
\end{figure}

Figures~\ref{fig3}(a) and 3(b) show the detailed momentum dependence of the $2\omega$ and $\omega$ peaks in $I_\mathrm{s}$ at $\omega_1=6t$ and $9t$, respectively.  The former, plotted as functions of $E_\mathrm{kin}-\omega_1-\omega_2$, can be compared to the occupied side of the single-particle spectral function $A(\mathbf{k},\omega)$ at half filling~\cite{Tsutsui} in Fig.~\ref{fig3}(c), while the latter can be compared to the unoccupied side.  As expected, $I_\mathrm{s}$ at $\omega_1=6t$ shows momentum dependence globally similar to that of $A(\mathbf{k},\omega)$ in the energy region of $-8<\omega/t<0$, although the details of spectral shape, for example, at $\mathbf{k}=(0,0)$ are different. We note that the spectral weight below $E_\mathrm{kin}-\omega_1-\omega_2=-8t$ is due to one-photon photoemission keeping an electron in the UHB. In $I_\mathrm{s}$ at $\omega_1=9t$, despite the broad spectral-weight distributions and additional weights at around $E_\mathrm{kin}-\omega_2=3t$ in the $\mathbf{k}=(0,0)$, $(\pi/2,0)$, and $(0,\pi/2)$ spectra, the momentum dependence show a similarity to that of $A(\mathbf{k},\omega)$ for $\omega>0$: From $(\pi,\pi)$ to $(\pi,0)$ and $(0,\pi)$, the spectra show a dispersive feature qualitatively consistent with the dispersion in the UHB.  However, because of diffusive features in the spectra, it seems to be difficult to identify the location of the bottom of the UHB exactly from the $\omega$-peak analysis. 

Next we consider the case where the energy of the pump photon is tuned to an excitation from the NB oxygen band to UHB [see Fig.~\ref{fig1}(b)].  In order to simplify the problem, we construct the Wannier orbitals centered at the copper and oxygen sites according to Ref.~1, and rewrite a current operator between copper and oxygen sites by using the Wannier orbitals.  After all, we obtain an expression of the cuurent operator between the occupied NB band and unoccupied UHB: $j_x=\sum_{\mathbf{k},\sigma} \beta_\mathbf{k} c_{\mathbf{k},\sigma}^\dagger b_{\mathbf{k},\sigma}$, where $b_{\mathbf{k},\sigma}$ is the annihilation operator of the NB state, and $\beta_\mathbf{k}=C \cos (k_x/2) \sin (k_y/2) / \sqrt{1-(\cos k_x +\cos k_y)/2}$ with $C=2d_\mathrm{Cu-O}T_{pd}$, $d_\mathrm{Cu-O}$ and $T_{pd}$ being the distance and hopping amplitude between neighboring Cu and O, respectively.  In the present work, we assume that (i) the dispersion of the NB band is negligible~\cite{Nonbonding} and (ii) there is no interaction between the hole left in the NB band and electrons in both LHB and UHB.  Under these assumptions, we can neglect $b_{\mathbf{k},\sigma}$ in $j_x$.  Thus, the dipole transition process results in an electron-addition process with the momentum-dependent factor $\beta_\mathbf{k}$.  We also take the coefficient $C$ to be unity for simplicity.  In addition to these simplifications, we use the $t$-$t'$-$t''$-$J$ model instead of Eq.~(\ref{HtU}), since only UHB is necessary in this process.  We note that the $t$-$t'$-$t''$-$J$ model gives almost the same results as those of the $t$-$t'$-$t''$-$U$ model when $U$ is large enough to satisfy $J=4t^2/U$.  The chemical potential of the half-filled $t'$-$t'$-$t''$-$J$ model is determined in order for LHB and UHB to be separated by $U$.~\cite{Tohyama}  

\begin{figure}
\begin{center}
\includegraphics[width=7.5cm]{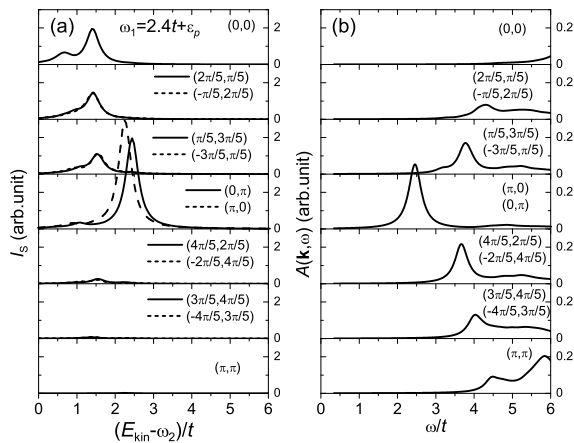}
%\vskip4cm
\caption{\label{fig4}
(a) Angle-resolved two-photon photoemmision spectra $I_\mathrm{s}$ from the simultaneous process in the case that the pump photon excite an electron from the NB band, obtained by using a $\sqrt{20}\times\sqrt{20}$ $t$-$t'$-$t''$-$J$ cluster.  The energy of the pump photon is tuned to the energy deference between the NB band and the bottom of the UHB with momentum $(\pi,0)$ and $(0,\pi)$ as shown in the electron-addition spectral function $A(\mathbf{k},\omega)$ in (b).}
\end{center}
\end{figure}

In Fig.~\ref{fig4}, we show $I_\mathrm{s}$ obtained by tuning $\omega_1$ to an excitation energy from the localized NB band to the bottom of UHB located at $(\pi,0)$ and $(0,\pi)$.  We note that only the $(0,\pi)$ state is occupied by the excitation according to the momentum-dependent coefficient $\beta_\mathbf{k}$.  Here $\omega_1=\varepsilon_p+2.4t$ with the level of the NB oxygen band $\varepsilon_p$.  The highest-energy structure in $I_\mathrm{s}$ is located at $(0,\pi)$ and its energy is the same as that in $A(\mathbf{k},\omega)$.  This means that, by tuning the pump-photon energy to the bottom of the UHB, we can observe its position in the momentum and energy spaces.  If we increase $\omega_1$ to be tuned to a quasiparticle-peak position at $\mathbf{k}=(4\pi/5,2\pi/5)$ with $\omega=3.7t$ in Fig.~\ref{fig4}(b), the high-energy edge of the 2P-APRES spectra appears at $(4\pi/5,2\pi/5)$ with $E_\mathrm{kin}-\omega_2=3.7t$ (not shown).  Therefore, the dispersion of the UHB near $(\pi,0)$ and $(0,\pi)$ would be detectable by changing $\omega_1$, provided the dispersion of the NB band is negligible.~\cite{Nonbonding}  We note that $I_\mathrm{c}$ shows the same behaviors as those of $I_\mathrm{s}$ (not shown).

In addition to the $(0,\pi)$ structure, there are spectral weights at smaller momentum region with lower kinetic energies.  For example, a peak appears at $(0,0)$, separated from the $(0,\pi)$ peak by $\sim t$, which is not present in $A(\mathbf{k},\omega)$. Such additional structures come from the following reason.  In the process where the NB state is excited, the final state $\left|f\right\rangle$ belongs to the half-filled system whose low-lying excitations are of the spin wave.  Actually the $(0,0)$ peak exists at the eigenstate of the spin-wave excitation.   Since the peak at $(0,\pi)$ comes from the ground state of the half-filled system, the energy separation between the $(0,\pi)$ and $(0,0)$ peaks is the same as the spin-wave width between the momentum transfers $\mathbf{q}=(0,0)$ and $(0,\pi)$.  Accordingly the $\mathbf{q}=(\pi,\pi)$ spin-wave state contributes to the $(\pi,0)$ peak in Fig.~\ref{fig4}(a), which is thus expected to be degenerate with the $(0,\pi)$ peak in the thermodynamic limit.  After all, we can say that the additional states in AR-2PPES contain knowledge of the spin excitation at half filling.

Finally we comment on experimental conditions that would confirm the present theoretical results.  The most crucial point is whether one can get $E_\mathrm{kin}$ enough to reach to the momentum $(\pi,0)$ and $(0,\pi)$.  Here we note that the maximum of $E_\mathrm{kin}$ is given by $E_\mathrm{kin}^\mathrm{max}=E_\mathrm{gap}/2+\omega_2$, $E_\mathrm{gap}$ being the Mott-gap magnitude with approximately $4t$ according to Fig.~\ref{fig3}(c).  In the case of the excitation from LHB to UHB, $\omega_1$ should be around $9t$ from Figs.~\ref{fig2} and \ref{fig3}. If $\omega_2=\omega_1$, $E_\mathrm{kin}^\mathrm{max}=11t\sim 4$~eV.  In the case from the NB band, $\omega_1$ would be $9t+E_\mathrm{B}$, where $E_\mathrm{B}$ is the binding energy of ZRB and approximately 2~eV.  Thus, by assuming $\omega_2=\omega_1$, $E_\mathrm{kin}^\mathrm{max}\sim 6$~eV in this case.  On the other hand, the minimum value of $E_\mathrm{kin}$ necessary to reach $(\pi,0)$ can be estimated to be $\sim 6$~eV, by taking the lattice constant ($\sim 3$~\AA) and the work function ($\sim 4$~eV) into account.  This leads to the conclusion that, under the condition  $\omega_2=\omega_1$, the momentum $(\pi,0)$ cannot be reached for the excitation from LHB but can critically for the excitation from the NB band.  If we take $\omega_2>\omega_1$, the condition is relaxed and the possibility to observe the $(\pi,0)$ state is enhanced.

In summary, we have proposed that AR-2PPES is a promising technique to observe the location of the bottom of the UHB in 2D insulating cuprates.  When the pump photon is tuned to an excitation from the LHB to UHB, the bottom of the UHB is less clear because of diffusive spectral features.  On the other hand, when the photon energy is tuned to an excitation from NB oxygen band, we clearly see the bottom of the UHB.  In addition to this, additional spectra that are not present in inverse photoemission spectra emerge in AR-2PPES.  Their origin is attributed to the spin excitation expected at half filling.

We would like to thank Y. Sonoda and T. Munakata for enlightening discussions.  We am also grateful to D. L. Feng, S. Shin, and T. Takahashi for useful discussions.  This work was supported by NAREGI Nanoscience Project, CREST, and Grant-in-Aid for Scientific Research from the Ministry of Education, Culture, Sports, Science and Technology of Japan.


\begin{thebibliography}{99}
\bibitem{Nonbonding} Although no dispersion is assumed for the NB band, this is not necessary to observe the bottom of UHB at $(\pi,0)$. Rather an important condition is that the top of the NB band is near the $(\pi,0)$ region because the dipole transition with minimum-excitation energy occurs at the region.   A preliminary result of the spectral function for a realistic three-band Hubbard model including Cu3$d_{x^2-y^2}$ and O2$p_\sigma$ orbitals satisfies this condition.
\end{thebibliography}
\end{document}